A translation of Pierre Marx's 'Effets gravitationnels des champs électromagnétiques intenses'

Gravitational effects of intense electromagnetic fields

PIERRE MARX
102, les bois du cerf, 91450 Étiolles, France
marx.p@wanadoo.fr

ABSTRACT. In the presence of an electromagnetic (EM) field the space-time metrics are affected by the field potential. The principle of superposition, which usually rules EM fields, is no longer valid for the biggest fields and Maxwell's equations, for the free field, generally have a non-zero right-hand side. For weak potentials the metrics are very close to Euclidean metrics i.e. the gravitational field is negligible. Very high EM potentials and fields are necessary to create a measurable gravitational field and even higher to fail the principle of superposition. In this last case, the energy-impulsion tensor of the EM field has to be modified and the linearization of Einstein's equations is no longer possible.

Introduction

According to the theory of general relativity, any source of energy (matter or field) creates a gravitational field. This is the case of E.M. energy. But, compared with those of matter, the gravitational effects of the E.M. usual fields are extremely low and are far from being measurable as discussed below. This is probably why this question doesn't interest contemporary physics and seems to have been treated nowhere. However, we can't exclude that one day we know amplify or combine these effects in order to create a significant gravitational field on the macroscopic scale. If the experiments made by E. Podkletnov, M. Najjar and consort on superconducting disks, if proven evidence (which doesn't seem the case) could be interpreted that way.
It is in this perspective, perhaps illusory, that fits this article.

2. Gravitation and superposition principle

2.1. Validity of the superposition principle

Property of the electromagnetic field (E.M.) into the void to obey the superposition principle is an experimental fact. However, there is no physical system where the effect (by induction) is proportional to the cause (by field) whatever its intensity. There is reason to think, either that the superposition principle has not been validated for very high field, or that fields likely to invalidate the principle has never been observed or even that they are so intense that we don't know to produce them.
For these reasons, the classical theory of the E.M. field is based on the superposition principle which implies that the differential equations governing the field (the second group of Maxwell's equations) must be linear ([1], § 27). This fact, combined with the one of the potential ambiguity defines the Lagrangian as a quadratic form of the field components:

$$\Lambda = -\frac{\varepsilon_0}{4} F^{kl} F_{kl} = -\frac{\varepsilon_0}{4} \left( g^{ik} g^{jl} \right) F_{ij} F_{kl} \quad (2.1)$$

where the quantities, coefficients of the form, are data at the considered point. We obtain the Minkowski's equations of electrodynamics in space in Galilean coordinates ([1], § 30) or in curvilinear coordinates [1], § 90) associated, either with a simple change of coordinates without changing the metric, or to a gravitational field not due to an E.M. field[1].

Thus, the linearity of Maxwell's equations implies that the metric tensor doesn't depend on the E.M. field while general relativity provides the contrary. The reason is precisely that the gravitational effects are only significant for very intense E.M. fields. This suggests that, into the void, *the superposition principle applies to E.M. fields probably in a very wide domain but nevertheless finished.*

2.2. hypothesis

---

[1] The electric D and magnetic B inductions are then related to the fields E and H by the relations ([[1], § 90):

$$\boldsymbol{D} = \frac{\boldsymbol{E}}{\sqrt{g_{00}}} + \boldsymbol{H} \times \boldsymbol{g} \; ; \; \boldsymbol{B} = \frac{\boldsymbol{H}}{\sqrt{g_{00}}} + \boldsymbol{g} \times \boldsymbol{E}$$

where g is the three-dimensional vector of $g^{\alpha} = -g^{0\alpha}$ components.

To take into account the fact that the E.M. field generates a gravitational field ([1], § 95) and therefore that the metric is affected by the field, it must be added in the classical expression of the variation of the Lagrangian $\Lambda\sqrt{-g}$ for the free field:

$$\delta\left(\Lambda\sqrt{-g}\right) = \frac{\partial \Lambda\sqrt{-g}}{\partial A_{k,l}} \delta A_{k,l}$$

a term representing the variation of the metric coefficients, so be it:

$$\delta\left(\Lambda\sqrt{-g}\right) = \frac{\partial \Lambda\sqrt{-g}}{\partial g^{ij}} \delta g^{ij} + \frac{\partial \Lambda\sqrt{-g}}{\partial A_{k,l}} \delta A_{k,l} \quad \textbf{(2.2)}$$

To make the action stationary, it must connect the metric to E.M. field. To do this, we propose to approximate the expression above of that, generally, of the Lagrangian variation:

$$\delta\left(\Lambda\sqrt{-g}\right) = \frac{\partial \Lambda\sqrt{-g}}{\partial A_k} \delta A_k + \frac{\partial \Lambda\sqrt{-g}}{\partial A_{k,l}} \delta A_{k,l} \quad (2.3)$$

which suggests to pose: $\delta g^{ij} = \frac{\partial g^{ij}}{\partial A_k} \delta A_k$

Hence the hypothesis:

"In the presence of an electromagnetic field, the space-time metric depends on fields potentials ".

Under these conditions, the coefficients are functions of geometric coordinates, first via the generalized coordinates, the other due to any other forms of energy (i.e. other gravitational sources) or if the 4-space is plan and reported to curvilinear coordinates. So we pose:

$$g^{ij}\left(x^l\right) = g^{ij}\left[A_k\left(x^l\right), x^l\right] \quad (2.4)$$

From the standpoint of general relativity, we don't restrict the generality Einstein equations. The gravitational field, that is to say, the metric is defined by all forms of energy in presence. Merely, the part due to the E.M. field is, according to our hypothesis, defined by the potential field.

Justification

In the space configuration of a physical system, the kinetic energy may depend on generalized coordinates via the coefficients of this form ([1], § 5). This possibility can be applied to the Lagrangian of the E.M. field, quadratic form of generalized velocities (covariant components of the Faraday tensor):

$$\Lambda = -\frac{\varepsilon_0}{4} F^{kl} F_{kl} = -\frac{\varepsilon_0}{4} g^{ik} g^{jl} F_{ij} F_{kl} \quad (2.5)$$

This is not the case in classical theory where the coefficients $g^{ik} g^{jl}$ depend only on the geometric coordinates, that don't vary when varying the Lagrangian of the field.

From a formal point of view, this situation is similar to that which determines the trajectory of a free particle from the Lagrangian: $L = -mc\,ds/dt$ ([1], § 87) depending on whether one is in the plane space in Galilean coordinates or not. The difference relates to the position variables: potential for the E.M. field, geometric coordinates for the particle.

Finally, from a mathematical point of view, it is only a change of variables. The 10 independent functions $g^{ij}$ are for the part which falls within the E.M. field, functions of four generalized coordinates, which are functions of four geometric coordinates $x^l$.

2.3. Equivalence principle

The total derivatives of the metric coefficients with respect to geometric coordinates are written:

$$\frac{\partial g^{ij}}{\partial x^l} = A_{k,l} \left(\frac{\partial g^{ij}}{\partial A_k}\right)_{M \text{ donné}} + \left(\frac{\partial g^{ij}}{\partial x^l}\right)_{A \text{ donné}} \quad (2.6)$$

The first term represents the metric variation due to the 4-potential A at a given point, the second includes variation of the metric due to the presence of other forms of energy and that of the metric coefficients (to metric given) if the 4-space is reported in curvilinear coordinates:
- For low Potentials, it only remains the second term. In the absence of other sources of gravitation, the space is flat and it is always possible to relate to Galilean coordinates. Without other sources of gravitation, the space is flat and it is always possible to relate him to Galilean coordinates.
- High values of E.M. field potentials generate a gravitational field (first term).
*As regards the metric, the two terms have a similar role in accordance with the fundamental principle of general relativity.* In particular, it is always possible to cancel locally the gravitational field in an accelerated reference.

2.4. Maxwell's equations modified

It results from what precede, that in a point of coordinates $x^l$ given, the variation of the Lagrangian is:

$$\delta\left(\Lambda\sqrt{-g}\right) = \frac{\partial \Lambda\sqrt{-g}}{\partial g^{ij}} \frac{\partial g^{ij}}{\partial A_k} \delta A_k + \frac{\partial \Lambda\sqrt{-g}}{\partial A_{k,l}} \delta A_{k,l} \quad (2.7)$$

or by introducing the TEI E.M. field ([1], § 94):

$$\delta\left(\Lambda\sqrt{-g}\right) = \frac{1}{2}\sqrt{-g}\,\tau_{ij}\frac{\partial g^{ij}}{\partial A_k}\delta A_k + \frac{\partial \Lambda\sqrt{-g}}{\partial A_{k,l}}\delta A_{k,l} \quad (2.8)$$

Hence Maxwell's equations for the free field:

$$\varepsilon_0 D_l F^{kl} = \frac{1}{2}\tau_{ij}\frac{\partial g^{ij}}{\partial A_k} \quad (2.9)$$

Note: it follows that the derivatives $\partial g^{ij}/\partial A_k$ are the components of a tensor.

Thus, Maxwell's equations for the E.M. free field have generally a second member not zero, however, completely negligible for the usual fields.

<u>Density</u> of <u>electro gravitational current</u>

By analogy with Maxwell's equations for the field in the presence of charges: $\varepsilon_0 D_l F^{kl} = -\frac{1}{c} j^k$ where j is the 4- density of electric current, we put:

$$-\frac{1}{c}f^k = \frac{1}{2}\tau_{ij}\frac{\partial g^{ij}}{\partial A_k} \quad (2.10)$$

We propose to call the four-vector f: "4- density of electro gravitational current" or an abbreviation, "4-density of egc".

As the 4- density of electric current, the 4-density of egc satisfies the law of charge conservation:

$$D_k f^k = -c\varepsilon_0 D_k \left(D_l F^{kl}\right) \equiv 0 \quad (2.11)$$

*Discussion*

In the classical case, the E.M. field and the potential which it is derived are too weak to generate a significant gravitational field. So $f \approx 0$. We find the classical Maxwell equations which applies the superposition principle.

There are two other cases where the second member of equations (3) is null:
- The 4-potential is very high but almost uniform. Derivatives $A_{k,l}$ are null or very small so that the E.M. field and the TEI associated are close to zero. Although the quantities $\partial g^{ij}/\partial A_k$ are not null, the derivatives: $\frac{\partial g^{ij}}{\partial x^l} = \frac{\partial g^{ij}}{\partial A_k}\frac{\partial A_k}{\partial x^l}$ are null or very small so that there is no or little gravity,
- Neither $\partial g^{ij}/\partial A_k$ nor $\tau_{ij}$ aren't all null but their product is null. So there is EM field and gravitation.

3. Gravitational potentials for a low E.M. field

3.1. Preliminaries

*Continuity of the metric*

The gravitational potentials (metric coefficients) are assumed to be determined by the 4-potential A when it is very high, we must tend to the Euclidean metric (Minkowski space characterized by the absence of gravity) for small values of the latter (assuming that the E.M. field is the only form of energy in presence).

A first question is to know if we pass to the metric depending on A to the Euclidean metric on a continuous or discontinuous way:
- In the first case, the metric tends continuously towards the Euclidean metric when A tends to 0, so be it: $A \to 0 \Rightarrow g_{ij} \to g_{ij}^{(0)}$ where the $g_{ij}^{(0)}$ are the coefficients of the Euclidean metric in an arbitrary coordinate system.
- In the second case, there is a 4-potential $A^{(0)}$ for which the transition takes place:
$A^2 < A^{(0)2} \Rightarrow g_{ij} = g_{ij}^{(0)}$ ; $A^2 \geq A^{(0)2} \Rightarrow g_{ij} = g_{ij}(A_k)$ (3.1)

Similarly, Maxwell's equations with second member must give the classical equations without second member when the superposition principle applies. As before, two cases are possible:
- Or the second member continuously tends to 0 when A tends to 0. So, the TEI E.M. field is not null for low values of the field, it is $\partial g_{ij}/\partial A_k$ which tend to 0 when A tends to 0:

$$A \to 0 \Rightarrow \frac{\partial g_{ij}}{\partial A_k} \to 0 \quad (3.2)$$

- Or there is a 4-potential $A^{(0)}$ for which the transition takes place:

$$A < A^{(0)} \Rightarrow \frac{\partial g_{ij}}{\partial A_k} = 0 \ ; \ A \geq A^{(0)} \Rightarrow \frac{\partial g_{ij}}{\partial A_k} \neq 0 \quad (3.3)$$

The simplest hypothesis is that the Einstein equations apply whatever the level of momentum-energy involved. Simply the gravitational field generated can only be detected at extremely high levels.

It follows that the metric coefficients and their first derivatives with respect to the components of A must be continuous functions. Hence we have:

$$A \to 0 \Rightarrow \left| \begin{array}{l} g_{ij} \to g_{ij}^{(0)} \\ \dfrac{\partial g_{ij}}{\partial A_k} \to 0 \end{array} \right. \quad (3.4)$$

Einstein's equations

As the metric is based on the E.M. field potentials, it must provide the Einstein's equations for a pure E.M. field, namely ([1], § 95):

$$E_{ik} = R_{ik} - \frac{1}{2} g_{ik} R - g_{ik} \mathcal{L}' = \frac{8\pi G}{c^4} \tau_{ik} \quad (3;5)$$

Conversely, we must determine the metric by integrating these equations.

The cosmological constant, denoted here[2] $\mathcal{L}'$, may appear at the first member (term: $-g_{ik}\mathcal{L}'$) or at the second under the form: $+g_{ik}\mathcal{L}' = +g_{ik}\dfrac{c^4}{8\pi G}\mathcal{L}$, according to the interpretation that one makes (geometric or physical).

τ is the TEI E.M. field components:

$$\tau_{ik} = -\varepsilon_0 F_i{}^l F_{kl} - g_{ik} \Lambda \quad (3.6)$$

*thereby limiting the validity of the previous Einstein's equations to low E.M. fields.*

Indeed, the 4-divergence of τ is null (condition that must be satisfied the second member tensor of Einstein's equations) that if the 4-divergence of the Faraday tensor is null itself (Maxwell's equations without second member), which according to our hypothesis, is not the case for the very intense E.M. field (see § 3).

Proposed Approach

It appears difficult to integrate the Einstein's equations in the most general case. That's why we have restricted ourselves here to particular cases. We chose two amongst of the usual fields:
- The uniform electrostatic field, the easiest E.M. fields
- A plane progressive wave linearly polarized, basic element of a pure E.M. field representation.

For conventional fields, the gravitational field is extremely low so that we can consider the associated metric is almost Euclidean.

Drawing on the method used for gravitational waves ([1], § 102) where one is in the same case, we pose: $g_{ik} = g_{ik}^{(0)} + h_{ik}$ where the $g_{ik}^{(0)}$ are the metric tensor components in an orthonormal

---

[2] It is usually denoted by $\Lambda$. Here we note it $\mathcal{L}$, reserving the notation $\Lambda$ to the Lagrangian E.M. field

reference frame $\left(g_{00}^{(0)}=1\,;\,g_{\alpha\alpha}^{(0)}=-1\,;\,g_{ij}^{(0)}=0 \text{ pour } i\neq j\,;\,g^{(0)}=-1\right)$ and the $h_{ik}$ the small corrections that determine the gravitational field.

In making the change of variable:
$$\psi_k^l = h_k^l - \frac{1}{2}\delta_k^l h \Leftrightarrow h_k^l = \psi_k^l - \frac{1}{2}\delta_k^l \psi \quad (3.7)$$
the Einstein's tensor components (without cosmological constant) are written:
$$E_{ik} = \frac{1}{2}\frac{\partial}{\partial x_l}\left(\frac{\partial \psi_{il}}{\partial x^k} + \frac{\partial \psi_{kl}}{\partial x^i} - \frac{\partial \psi_{ik}}{\partial x^l} - g_{ik}\frac{\partial \psi_{lm}}{\partial x_m}\right) \quad (3.8)$$
Note: in the case of gravitational waves when the second member of Einstein's equations is null, the additional conditions: $\partial_l \psi_k^l = 0$ lead to the classical wave's equation: $\Box \psi_{ik} = 0$. But these conditions assume the choice of a particular coordinate system a priori different from the one which the TEI field is reported. Alpha
Although they are very similar because of the small quantities $\psi_i^k$, changes in components of the metric tensor are the same as those due to the E.M. field while the variation of the TEI field is second order because of the smallness of the amount $8\pi G/c^4$. It must therefore remain in the same coordinate system and to consider the quantities $\psi_i^k$ only as a convenient change of variables.

3.2. Case of uniform electrostatic field

If E is carried by the x-axis, the system depends only on the spatial coordinate $x = x^1$. Then:
$$E_{00} = \frac{1}{2}\frac{d^2(\psi_{00} - \psi_{11})}{dx^2}\,;\,E_{11} = 0\,;$$
$$E_{22} = \frac{1}{2}\frac{d^2(\psi_{11} + \psi_{22})}{dx^2}\,;\,E_{33} = \frac{1}{2}\frac{d^2(\psi_{11} + \psi_{33})}{dx^2}$$
$$E_{02} = \frac{1}{2}\frac{d^2\psi_{02}}{dx^2}\,;\,E_{03} = \frac{1}{2}\frac{d^2\psi_{03}}{dx^2}\,;\,E_{23} = \frac{1}{2}\frac{d^2\psi_{23}}{dx^2}$$
$$E_{01} = E_{12} = E_{13} = 0$$
Furthermore, the only nonzero components of the TEI field are:
$$\tau_{00} = -\tau_{11} = \tau_{22} = \tau_{33} = \frac{\varepsilon_0}{2}E^2 \quad (3.10)$$
The second equation $E_{11} = \frac{8\pi G}{c^4}\tau_{11}$ has no solution other solution than $E = 0$. We must therefore introduce a "cosmological" constant. By putting it at the second member, it comes:
$$0 = \frac{8\pi G}{c^4}\left(-\frac{\varepsilon_0}{2}E^2 + g_{11}\mathcal{L}\right) \Rightarrow \mathcal{L} = -\frac{\varepsilon_0}{2}E^2 \quad (3.12)$$
Einstein's equations are then written:
$$E_{ik} = \frac{8\pi G}{c^4}(\tau_{ik} - g_{ik}\mathcal{L}') = \frac{8\pi G}{c^4}\sigma_{ik} \quad (3.12)$$

$$\sigma_{ik} = \tau_{ik} - g_{ik}\mathcal{L}' \Rightarrow \sigma_{00} = \sigma_{11} = 0\,;\,\sigma_{22} = \sigma_{33} = \varepsilon_0 E^2 \quad (3.13)$$
Hence the six equations:

$$\frac{d^2(\psi_{00} - \psi_{11})}{dx^2} = 0 \Rightarrow \psi_{00} = \psi_{11}$$

$$\frac{d^2(\psi_{11} + \psi_{22})}{dx^2} = \frac{d^2(\psi_{11} + \psi_{33})}{dx^2} = \frac{16\pi\varepsilon_0 G}{c^4} E^2 \Rightarrow \psi_{22} = \psi_{33}$$

$$\frac{d^2\psi_{02}}{dx^2} = \frac{d^2\psi_{03}}{dx^2} = \frac{d^2\psi_{23}}{dx^2} = 0 \Rightarrow \psi_{02} = \psi_{03} = \psi_{23} = 0$$

The quantities: $\psi_{01}, \psi_{12}, \psi_{13}$ do not appear and therefore should be considered null (since only those determined by the field changes the metric). Hence: $i \neq k \Rightarrow \psi_{ik} = 0$. We deduce the quantities $h_{ik}$:

$$\frac{d^2 h_{00}}{dx^2} = \frac{16\pi\varepsilon_0 G}{c^4} E^2 \; ; \; h_{11} = \psi_{00} - \psi_{22} \; ; \; h_{22} = h_{33} = 0 \; ; \; i \neq k \Rightarrow h_{ik} = 0 \quad (3.14)$$

The amount $h_{11} = \psi_{00} - \psi_{22}$ is undetermined. We must therefore considered it as zero (see above) so that the only amount $h_{ik}$ non-zero is $h_{00}$. Then:

$$\frac{d^2 h_{00}}{dx^2} = \frac{16\pi\varepsilon_0 G}{c^4} E^2 \Rightarrow h_{00} = \frac{8\pi\varepsilon_0 G}{c^4} V^2 \quad (3.15)$$

where $V$ is the electrostatic potential which derives the field $E$. Introducing the Plank potential: $V_p = \sqrt{\dfrac{c^4}{4\pi\varepsilon_0 G}}$, we finally obtain:

$$g_{00} = 1 + \frac{2}{V_p^2} V^2 \quad (3.16)$$

Note: The spatial coordinates remain orthonormal. The temporal coordinate is no longer normalized. but remain orthogonal to the other since $h_{0\alpha} = \psi_{0\alpha} = 0$.

3.3. Case of the plane progressive wave

- Here the field depends on both coordinates $x = x^1$ and $t = x^0/c$. It is a function of the variable $w = x - ct$, x being the propagation direction.
- We consider a wave linearly polarized (not necessarily monochromatic): induction $B$ defines the y-axis and E, the z axis.
- The potential vector A is carried by the z-axis (to simplify the notation, we denote $A$ the component $A_z$). The other components of the four-vector $A$ are null:

$$A_0 = V = 0 \; ; \; A_1 = A_2 = 0 \; ; \; A_3 = -cA_z = -cA \quad (3.17)$$

Furthermore, the only non-zero components are:
- for the Faraday tensor:

$$F_{03} = E_z = E = -\frac{\partial A}{\partial t} \; ; \; F_{13} = cB_y = cB = -c\frac{\partial A}{\partial x} \quad (3.18)$$

- For the TEI field: $\tau_{00} = -\tau_{01} = \tau_{11} = \varepsilon_0 E^2 \quad (3.19)$

Finally:

$$A = A(x - ct) \Rightarrow \frac{\partial}{\partial x} = -\frac{1}{c}\frac{\partial}{\partial t}$$

$$\Rightarrow \begin{vmatrix} \dfrac{\partial A}{\partial x} = -\dfrac{1}{c}\dfrac{\partial A}{\partial t} \\ \dfrac{\partial E}{\partial x} = -\dfrac{1}{c}\dfrac{\partial E}{\partial t} \end{vmatrix} \Rightarrow E = -cB \Rightarrow F_{03} = -F_{13}$$

We deduce the 5 equations:

$$\frac{d^2(\psi_{00}-\psi_{11})}{dw^2} = \frac{16\pi\varepsilon_0 G}{c^4}\boldsymbol{E}^2 = \frac{4}{V_p^2}\boldsymbol{E}^2 \ ; \ \psi_{01}=0 \ ; \ \psi_{02}+\psi_{12}=0 \ ; \ \psi_{03}+\psi_{13}=0 \ ; \ \psi_{23}=0$$

Hence: $\psi = \psi_{00} - (\psi_{11}+\psi_{22}+\psi_{33}) \Rightarrow \dfrac{d^2\psi}{dw^2} = \dfrac{4}{V_p^2}\boldsymbol{E}^2 - \dfrac{d^2(\psi_{22}+\psi_{33})}{dw^2}$. Then:

$$h_{00}-h_{11} = \left(\psi_{00}-\frac{1}{2}\psi\right) - \left(\psi_{11}+\frac{1}{2}\psi\right) \Rightarrow \frac{d^2(h_{00}-h_{11})}{dw^2} = \frac{d^2(\psi_{22}+\psi_{33})}{dw^2} \Rightarrow h_{00}-h_{11} = \psi_{22}+\psi_{33}$$

$$\left.\begin{array}{l} h_{22} = \psi_{22}+\dfrac{1}{2}\psi \Rightarrow \dfrac{d^2 h_{22}}{dw^2} = \dfrac{2}{V_p^2}\boldsymbol{E}^2 + \dfrac{1}{2}\dfrac{d^2(\psi_{22}-\psi_{33})}{dw^2} \\ h_{33} = \psi_{33}+\dfrac{1}{2}\psi \Rightarrow \dfrac{d^2 h_{33}}{dw^2} = \dfrac{2}{V_p^2}\boldsymbol{E}^2 - \dfrac{1}{2}\dfrac{d^2(\psi_{22}-\psi_{33})}{dw^2} \end{array}\right| \Rightarrow \dfrac{d^2(h_{22}+h_{33})}{dw^2} = \dfrac{4}{V_p^2}\boldsymbol{E}^2$$

$h_{01} = \psi_{01} = 0$ ; $h_{02}+h_{12} = \psi_{02}+\psi_{12} = 0$ ; $h_{03}+h_{13} = \psi_{03}+\psi_{13} = 0$ ; $h_{23}=\psi_{23}=0$

The amount $h_{00}-h_{11} = \psi_{22}+\psi_{33}$ is not determined by the field. Like $h_{00}$ and $h_{11}$ do not appear in any other equation, we must consider them as null, that is to say, not affecting the Galilean components. Hence $g_{00} = g_{00}^{(0)} = 1$ and $g_{11} = g_{11}^{(0)} = -1$.

The same reasoning applies to the quantities $h_{02}$ and $h_{12}$ and for $h_{03}$ and $h_{13}$ which only the sums are determined (equal to 0). We therefore have $g_{ik} = g_{ik}^{(0)} = 0$ for $i \neq k$.

In order to determine $h_{22}$ and $h_{33}$ which we know that the sum, we can see, by comparison with the previous case, that the diagonal component of the metric tensor affected by the field seems to correspond to the coordinate associated with the four-potential, here $x^3$, in which case we have:

$h_{22} = 0 \Rightarrow g_{22} = g_{22}^{(0)} = -1$

Then:

$$\frac{d^2 h_{33}}{dw^2} = \frac{4}{V_p^2}\boldsymbol{E}^2 \Rightarrow h_{33} = \frac{2}{V_p^2}c^2\boldsymbol{A}^2$$

Hence:

$g_{33} = -1 + \dfrac{2}{V_p^2}c^2\boldsymbol{A}^2$   (3-21)

3.4. Comparison

Metric

|  | ES field | Plane wave |
|---|---|---|
| Nonzero component of the 4-potential A | $A_0 = V$ | $A_3 = -c\boldsymbol{A}$ |
| 4-potential vector norm | $A^2 = V^2$ | $A^2 = -c^2\boldsymbol{A}^2$ |
| Components of the metric tensor affected by the E.M. field | $g_{00} = 1 + \dfrac{2}{V_p^2}A^2$ | $g_{33} = -1 - \dfrac{2}{V_p^2}A^2$ |

It is found that the diagonal component of the metric tensor affected by the field consists of the non-zero potential. We therefore have for these two particular cases:

$g_{ii} = g_{ii}^{(0)}\left(1 + \dfrac{2}{V_p^2}A^2\right)$   (3.22)

The orthogonality of the coordinates is preserved, the volume element affine being weighted by the g-density: $\sqrt{-g} = 1 + \dfrac{A^2}{V_P^2}$.

Cosmological constant

Unlike the case of the electrostatic field for which it is essential, the additional term, called "cosmological constant" for lack of better is not necessary in the case of the wave. We can also say, in this case, it is null or even that it represents the Lagrangian value $\Lambda$ (extremal constant value, the Maxwell equations being satisfied) and pose:

$\mathcal{L}' = \dfrac{8\pi G}{c^4} \Lambda$   (3.23)

Indeed, we have, for the two considered cases:

|  | ES field | Plane wave |
|---|---|---|
| Cosmological constant $\mathcal{L}' = \dfrac{c^4}{8\pi G}\mathcal{L}$ | $\mathcal{L} = \dfrac{\varepsilon_0}{2}E^2$ | $\mathcal{L} = 0$ |
| Lagrangian $\Lambda$ | $\Lambda = \dfrac{\varepsilon_0}{2}E^2$ | $\Lambda = \dfrac{\varepsilon_0}{2}\left(E^2 - c^2 B^2\right) = 0$ |

Under these conditions, the Einstein equations would be written:

$E_{ik} = \dfrac{8\pi G}{c^4} \sigma_{ik}$ ; $\sigma_{ik} = \tau_{ik} - g_{ik}\Lambda$   (3.24)

But this supposes that we have $\partial_k \Lambda = 0$, which is not the case in general.

That be, we found no essential purpose to this constant. We may simply observe that it would be include in the TEI field if we derivated relative to $g^{ik}$ the quantity $-\Lambda g$ and no $\Lambda\sqrt{-g}$.

3.5. Mechanical effects induced

The 4-acceleration of a particle in a gravitational field has for components ([1], § 87):

$\dfrac{du_i}{ds} = \dfrac{1}{2}\dfrac{\partial g_{kl}}{\partial x^i} u^k u^l$   (3.25)

In the case of the electrostatic field above, the only not zero component is:

$\dfrac{du_1}{ds} = \dfrac{1}{2}\dfrac{dg_{00}}{dx^1}\left(u^0\right)^2 \approx \dfrac{1}{2}\dfrac{dg_{00}}{dV}\dfrac{dV}{dx}$,

hence:

$\ddot{x} = \dfrac{2c^2}{V_P^2} E V = -\dfrac{c^2}{V_P^2}\dfrac{dV^2}{dx}$   (3.26)

For the plane progressive wave:

$\dfrac{du_i}{ds} = \dfrac{1}{2}\dfrac{\partial g_{33}}{\partial x^i}\left(u^3\right)^2 = \dfrac{1}{2}\dfrac{dg_{33}}{dA_3}\dfrac{\partial A_3}{\partial x^i}\left(u^3\right)^2$   (3.27)

the two not zero components are expressed as follows:

$$\frac{du_1}{ds} = -\frac{du_0}{ds} = -\frac{2c}{V_p^2} AE \left(\frac{v_z}{c}\right)^2 \quad (3.28)$$

Potentials and fields classification

These results suggest potentials and E.M. fields classification according to the product value of the potential by the field. In the case of ES field, the ratio $V_p^2/c^2$ of the product $EV$ to the gravitational acceleration is of the order of $10^{37}$. Thus:
- With an RFP of 1MV and a field of 3 MV / m, values[3] that can be qualified "intense", the gravitational acceleration is about $10^{-25}$ m/s², obviously not measurable[4],
- at the proton radius $(\approx 0,4.10^{-15}$ m$)$, the potential values 3.6 MV and the field: $9.10^{21}$ V/m, either $EV \approx 3.10^{27}$ V²/m. Hence an acceleration of the order of $10^{-10}$ m/s².
- to obtain an acceleration of $1$ m/s², a product is necessary $EV \approx 10^{38}$ V²/m, either on 1m, a DDP of $10^{19}$ volts.

*Such levels that we can characterize as "very intense" and that now seem largely unattainable could locally modify the usual gravity levels.*

4. Case of an extremely intense E.M. field

In what precedes, even the fields described as "very intense" affect only a little the Euclidean metric. The ceg density is negligible and the superposition principle remains entirely valid.
This situation changes drastically for fields deriving from commensurate potential with the Planck potential and that we agree to describe as "extremely intense":
- Firstly, the cge density can no longer be neglected,
- Secondly, the metric can no longer be considered as a small perturbation of the Euclidean metric.
These two points have the consequence of modifying the Einstein's equations and the metric determination.

---

[3] Performance of the best current electrostatic machines (Felici generator).

[4] The accuracy of the best current accelerometers is of $10^{-11}$ à $10^{-12}$ m/s² (eg: ONERA gradiometer for the GOCE mission of ESA).

## 4.1 Stress-Energy Tensor

### 4 - Divergence of the classical TEI

In the presence of charges (4- density of electric current j), 4-divergence of the TEI of a classic EM field is not null ([1], § 33):

$$D_l \tau_i^l = -\frac{1}{c} j^k F_{ik} \quad (4.1)$$

Given the formal equivalence of the CEG 4-density f, we can translate these relations to the case of the very intense free field and denote[5]:

$$D_l \tau_i^l = -\frac{1}{c} f^k F_{ik} \quad (4.2)$$

Thus, unlike the case of classical free field, *the 4-divergence of the tensor τ for the free EM field very intense is not null.*

In the classical theory of E.M. field in presence of charges, to the field tensor τ is necessarily adds the TEI charged matter[6]: $T_i^l = \mu c^2 u_i u^l \frac{ds}{dt}$ (**Erreur ! Source du renvoi introuvable.**, §33). Taking into account the laws governing the movement of charges in the field, the sum of 4-divergence of two tensors is null, a necessary condition to form the second member of Einstein's equations.

No such thing like in the case of EM free field extremely intense. It follows that, *in this case, the tensor τ can't yet be regarded as the TEI free field*. Thus, the concern is to know if a tensor exists that may play this role.

### TEI proposed

A simple transformation (see Annex) allows noting the second member of (4.2) under the form:

$$-\frac{1}{c} f^l F_{il} = -D_l \pi_i^l + \frac{1}{c} A^l s_{il} \quad (4.3)$$

In which π denotes the symmetric tensor components:

$$\pi_{ik} = -\frac{1}{c}\left(A_i f_k + A_k f_i - g_{ik} A_l f^l\right) \quad (4.4)$$

And s, the exterior derivative of f:

$$s_{ik} = -\frac{1}{c}(\partial_i f_k - \partial_k f_i) \quad (4.5)$$

Then:

---

[5] A more rigorous demonstration is given in Annex.
[6] According to A. Lichnerowicz, it should be cautious regarding the introduction into Einstein's equations of TEI's EM field, which is the result of restricted relativity and is alien to the principles underlying general relativity. This simple addition of the TEI of the field and matter is, in his view, an approximation that can only be provisional ([3], Book I, Ch.1, § II-9 and introduction to Book II).

$$D_l(\tau_i^l + \pi_i^l) = \frac{1}{c} A^l s_{il} \quad (4.6)$$

The tensor $\tau + \pi$ answers the question if the four-vector $\omega$ of components:

$$\omega_i = \frac{1}{c} A^l s_{il} \quad (4.7)$$

Is identically null.

*Another approach*

In § 1.2, we compared the case of the intense EM field with that of a free particle placed in a gravitational field. But at low velocities in front of c, the particle trajectory may as well be determined by the action of the Newtonian potential φ in Euclidean space. (**Erreur ! Source du renvoi introuvable.**, §87). Show that we can, in some way, find this double formulation for the EM field intense, due to the formal equivalence of 4- densities of the current and the CEG.

In the case of a classical EM field in the presence of charges, we add to the Lagrangian of free field (representing his "kinetic" energy T) the term: $-\frac{1}{c} A_k j^k$ who acts as a potential energy: -U ($-\frac{1}{c} j$ Being the "4-force"). By analogy, we write the Lagrangian of the intense field in the form:

$$\Lambda = T - U = -\frac{\varepsilon_0}{4} F^{kl} F_{kl} - \frac{1}{c} A_k f^k \quad (4.7)$$

The correspondent TEI is obtained by deriving $\Lambda\sqrt{-g}$ with respect to the contravariant components of the metric tensor ([1], § 87). The first term gives the usual tensor $\tau$. For the second, we can write:

$$\frac{\partial}{\partial g^{ik}}\left[\sqrt{-g}\left(-\frac{1}{c} A_l j^l\right)\right] = \frac{\partial}{\partial g^{ik}}\left[-\frac{1}{2c}\sqrt{-g}\, g^{lm}(A_l j_m + A_m j_l)\right]$$
$$= -\frac{1}{2c}\sqrt{-g}\left(A_k j_i + A_i j_k - g_{ik} A_l j^l\right) = \frac{1}{2}\sqrt{-g}\,\pi_{ik}$$

## 4.2 Einstein's equations modified

Under these conditions, the Einstein's equations must be written (assuming the "cosmological" constant contained in E):

$$E_{ik} = \frac{8\pi G}{c^4}(\tau_{ik} + \pi_{ik}) \quad (4.8)$$

Note: In the absence of "cosmological" constant and unlike the low EM field, the scalar curvature of 4-space is not null when the field is extremely intense:

$$R = -\frac{16\pi G}{c^5} A_k f^k \quad (4.9)$$

*Thus, subject to the nullity of the four-vector ω, the TEI of the free EM field appearing in the second member of Einstein's equations is generally the sum of two tensors τ and π. The latter is however completely negligible at the usual field.*

## *Verification*

Let's see what it's about in the particular case of the electrostatic field studied previously. We found the metric (3.16) by integrating the equation (3.15). But **E** cannot be uniform only if the second member of the Maxwell's equations is null, legitimate hypothesis in the case of a usual field. But, strictly speaking, we have:

$$E_{00} = \frac{1}{2} \frac{d^2 g_{00}}{dx^2} = \frac{2}{V_p^2} \left( \boldsymbol{E}^2 - V \frac{d\boldsymbol{E}}{dx} \right) \quad (4.10)$$

It's now the tensor π of components (4.5). In the present case, only their diagonal components are not null and we have:

$$\pi_{ii} = -\frac{1}{c} A_0 f_0 \; \forall i \quad (4.11)$$

$\varepsilon_0 D_l F^{kl} = -\frac{1}{c} f^k \Rightarrow f_0 = c\varepsilon_0 \frac{d\boldsymbol{E}}{dx}$ hence: $\pi_{00} = -\varepsilon_0 V \frac{d\boldsymbol{E}}{dx}$ that is well, after multiplication by $8\pi G/c^4$, the quantity to add to the second member of Einstein's equation if we do not do the approximation.

<u>Note:</u> The Maxwell's equations (2.9) give with the metric (3.16): $\frac{d\boldsymbol{E}}{dx} = \frac{1}{V_p^2} \boldsymbol{E}^2 V$. For d.d.p. of 1MV and a field of 3 MV/m (values taken previously for the mechanical effects), we get: $\frac{d\boldsymbol{E}}{dx} \approx 10^{-35}$ V.m$^{-2}$, Which justifies the hypothesis *a posteriori*.

**4.3 Metric**

The metric previously established (in two particular cases) for potentials up to "very intense" is not *a priori* be extrapolated to potentials "extremely intense". At most it should connect to it.

In the other way, it is reasonable to suppose that the potential EM field and/or the EM field itself, either reach an unsurpassable physical limit, or no longer suitable as a representation of physical reality[7].

We can illustrate this point by considering, for example, a metric of components ([1], § 92, Problem 2):

$$g_{ii} = g_{ii}^{(0)} \exp\left( \frac{2}{V_p^2} A^2 \right); \; g_{ik} = 0 \; \text{pour} \; i \neq k \quad (4.12)$$

which connects to the previous one for the "very intense" fields and that its limit of validity (or that of the induced gravitational field) come either from a maximal 4-potential and /or from limit EM field, whose values are very high[8].

---

[7] Following the example of metal bar, lying under the effect of a tensile force. As it is weak, the relative elongation is proportional. We are in the elastic domain which applies the principle of superposition. If we exceed a certain threshold, called elastic limit, elongation ceases to be proportional to the force exerted. Finally, beyond a certain elongation the bar breaks.

[8] For example, the potential of Planck $V_p \approx 10^{27}$ volts or for the field: $10^{16}$ V / m, The one created by an electron (charge $e = 1,6.10^{19}$ C ) at a distance equal to its radius quantum: $\bar{r} \approx \hbar / m_e c \approx 3,9.10^{-13}$ m .

Given we limit, in this article, to the "very intense" fields, sufficient to achieve levels of usual gravitation, we have not been more far in search of valid metric for the "very intense" fields.

For fields only "very intense", we tried to find a metric inspired by the form (3.22) but without success. For example, the metric:

$$g_{ik} = g_{ik}^{(0)} + \frac{4}{V_p^2}\left(A_i A_k - \frac{1}{2} g_{ik}^{(0)} A^2\right) \quad (4.13)$$

give the tensor $\tau + \pi$ plus additional terms that seem impossible to remove.

## Conclusions and perspectives

*Experimental validation*

The hypothesis that the metric of space-time may depend on the potential of the EM field is based, for the moment, than on arguments of principle. It is not experimentally verified today, at least to our knowledge. This can be done with a very high EM fields that we can possibly produce with high power lasers or among superconducting materials.

*Unification of the EM and gravitation*

The hypothesis builds a bridge between restricted relativity, natural setting of the classical EM field and general relativity, fundamental theory of gravitation. The EM field no longer appears as an exogenous factor which the TEI must be added to the second member of Einstein's equations, but, on the contrary, as one of its natural components. Viewed in this light, the proposed theory could help to unify the electromagnetism and gravitation, at least in classical settings[9].

However, this unification can only be partial until, considering only the EM field pure, it ignores the density of electric current, which remains, as says A. Lichnerowicz, "a foreign concept to the field" ([3], Introduction to Book II). For this reason, it is tempting to identify the 4- density of electric current *j* and electro gravitational *f*. The electric charge so appear as a consequence of the nonlinearity of the EM field and not as its source. But although there are formal elements in favor of this hypothesis, it is currently only an idea, without sufficient basis at this stage and that, in any event, beyond the scope of this article.

*An electromagnetic "ether"?*

Sources of gravitation, the potentials of the EM field, appear as a physical reality as well as matter. This status extends beyond their formal role of generalized coordinates. The EM field that is derived from it has, therefore, own existence, that recognizes the classical theory of EM waves (solutions of Maxwell's equations from the field without charges). Should we, so far, define them unambiguously? It is not obvious. An "EM space", reported to a system of coordinates $A_k$ of arbitrary origin, could be replaced by the geometric space, the distances and durations being defined from EM variables[10]. So reborn EM "ether" prepaid, this time, the contradictions of pre-relativistic[11] ether mechanistic.

---

[9] All attempts to unify electromagnetism and gravitation in classical settings failed. Today, this research is done in quantum (superstring theory and loop quantum gravity). Yet, formally, there is no objection to a unification in the classical framework, at least at the macroscopic scale ([3], Introduction to Book II).

[10] $[L]=[V][E]^{-1} ; [T]=[A][E]^{-1}$

[11] A development of General Relativity will lead Einstein to retrace his designs in 1905, as he explains in the conference who he delivered at the University of Leiden, May 5, 1920 ("Ether and the theory of General Relativity ").

## *Validity of Einstein's equations*

During the previous developments, appeared two points that are unclear:
- The "cosmological" constant that it had to be introduced in the special case of uniform electrostatic field, seems very artificial and its name, inappropriate (hence the quotes). But, in this specific case, we have not found any other way to form the Einstein's equations.
- The TEI proposed for the extremely intense field assumes that the 4-vector ω is null. Otherwise, we don't know how to write the Einstein's equations for the free field. We looked for a physical meaning of ω, for example such as a conservation law, but we didn't find any one.

- ***Control of gravitation***

Finally, if it were possible to create sufficient EM fields, the gravitational field that would be created would be opposed to a field of inertia or a gravitational field not caused by EM 4-potential and, therefore, allowing *local control of gravitation*. It is this dream that may seem foolish that motivated this approach and led to propose this theory. *It is, in fact, his real goal.*

# *Annex*

## TEI of the EM field extremely intense

We establish here, rigorously, the expression of TEI of the EM field extremely intense derived therefore, as in the classical case, the Lagrangian of the field relative to the spatial coordinates:

$$\frac{\partial \Lambda \sqrt{-g}}{\partial x^i} = \frac{\partial \Lambda \sqrt{-g}}{\partial A_{k,l}} \frac{\partial A_{k,l}}{\partial x^i} + \frac{\partial \Lambda \sqrt{-g}}{\partial g^{kl}} \frac{\partial g^{kl}}{\partial x^i}$$

Here, the coefficients of the metric involved in the Lagrangian of the field:

$$\Lambda_{\text{ch}} \sqrt{-g} = -\frac{\varepsilon_0}{4} \left( \sqrt{-g} \, g^{ik} g^{jl} \right) F_{ij} F_{kl}$$

depend on the coordinates geometry, first through the $A_k$ (gravitational field caused by 4-potential), on the other hand directly:

$$\frac{\partial g^{kl}}{\partial x^i} = A_{j,i} \left( \frac{\partial g^{kl}}{\partial A_j} \right)_{\text{M donné}} + \left( \frac{\partial g^{kl}}{\partial x^i} \right)_{A \text{ donné}}$$

The total derivative of the Lagrangian with respect to geometric coordinates are written as:

$$\frac{\partial \Lambda_{\text{ch}} \sqrt{-g}}{\partial x^i} = \frac{\partial \Lambda_{\text{ch}} \sqrt{-g}}{\partial A_{k,l}} \frac{\partial A_{k,l}}{\partial x^i} + \frac{\partial \Lambda_{\text{ch}} \sqrt{-g}}{\partial g^{kl}} \frac{\partial g^{kl}}{\partial A_j} \bigg|_{\text{M}} A_{j,i} + \frac{\partial \Lambda_{\text{ch}} \sqrt{-g}}{\partial g^{kl}} \frac{\partial g^{kl}}{\partial x^i} \bigg|_{A}$$

Or:

$$\frac{\partial \Lambda_{\text{ch}} \sqrt{-g}}{\partial g^{lm}} \frac{\partial g^{lm}}{\partial A_k} \bigg|_{\text{M}} = \frac{\partial \Lambda_{\text{ch}} \sqrt{-g}}{\partial A_k}$$

Given that the Maxwell's equations are verified, we get:

$$\frac{\partial \Lambda_{\text{ch}} \sqrt{-g}}{\partial x^i} = \frac{\partial}{\partial x^l} \left( \frac{\partial \Lambda_{\text{ch}} \sqrt{-g}}{\partial A_{k,l}} A_{k,i} \right) + \frac{1}{2} \sqrt{-g} \, \tau_{lm} \frac{\partial g^{lm}}{\partial x^i} + \frac{1}{c} \sqrt{-g} \, f^k A_{k,i}$$

Proceeding as above, we get:

$$\frac{1}{\sqrt{-g}} \frac{\partial}{\partial x^l} \left[ \sqrt{-g} \left( -\varepsilon_0 F^{kl} F_{ki} - \frac{1}{c} f^l A_i - \delta_i^l \Lambda_{\text{ch}} \right) \right] + \frac{1}{2} \tau_{lm} \frac{\partial g^{lm}}{\partial x^i} + \frac{1}{c} f^k A_{k,i} = 0$$

Where the relation (4.2): $\qquad D_l \tau_i^l = -\frac{1}{c} f^k F_{ik}$

The second member may develop as follows:

$$f^l F_{il} = f^l (D_i A_l - D_l A_i) = D_i (A_l f^l) - A_l D_i f^l - D_l (A_i f^l)$$
$$= D_i (A_l f^l) - A_l D_i f^l - D_l (A_i f^l) + \left[ -D_l (A^l f_i) + D_l (A^l f_i) \right]$$
$$= -D_l (A_i f^l + A^l f_i - \delta_i^l A_k f^k) - A^l (D_i f_l - D_l f_i)$$
$$= -D_l (A_i f^l + A^l f_i - \delta_i^l A_k f^k) - A^l s_{il}$$

Where the relation (4.6):

$$D_l (\tau_i^l + \pi_i^l) = \frac{1}{c} A^l s_{il}$$

with (4.4):

$$\pi_{ik} = -\frac{1}{c} (A_i f_k + A_k f_i - g_{ik} A_l f^l)$$

## *Main abbreviations and notations*

| EM | electromagnetism (tic) |
|---|---|
| $\Lambda$ | (density of the) Lagrangian EM field |
| $d\Omega$ | 4-volume elementary affine |
| $A$ | Vector four-potential of the EM field. |
| $V$ | scalar potential |
| $\mathbf{A}$ | vector potential |
| $F$ | Faraday tensor |
| $\mathbf{E}$ | electric field |
| $\mathbf{B}$ | magnetic induction |
| $j$ | four-vector current |
| $g_{ij}$ | metric tensor (components) |
| $g$ | determinant of the metric tensor |
| $R_{ik}$ | Ricci tensor (components) |
| $R$ | scalar curvature |
| $E$ | Einstein tensor |
| $\mathcal{L}$ | cosmological constant |
| TEI | Stress–energy tensor |
| $\tau$ | TEI of the EM field |
| $c$ | speed of light in vacuum |
| $G$ | constant of gravitation |
| $\varepsilon_0$ | permittivity of free space |
| $V_P$ | potential of Planck |

* This article is based almost exclusively on the classical relativistic theory of electromagnetic and gravitational fields, as outlined in the book by Landau and Lifshitz [1]. This is why it is <u>constantly</u> referred to this book. For this reason, we use, like in this book and contrary to widespread use, the Latin letters for the four-dimensional component indices and Greek letters for those three-dimensional components.